# BEYOND DETECTION: REDESIGNING ASSESSMENT AND GOVERNANCE OF GENERATIVE AI AT THE UNIVERSIDAD POLITÉCNICA DE MADRID (UPM)

**J. Díaz, S. Linio, F. Pescador, D. Martin-Fabiani**

*Universidad Politécnica de Madrid (SPAIN)*

## Abstract

Universities have responded to generative artificial intelligence (GenAI) in noticeably different ways, both internationally and within Spain. So far, the dominant reaction has been defensive, this is, most institutions frame the debate around AI detection, plagiarism, academic integrity and a presumed drop in student effort, prioritizing basic training for academic staff over students. Other group of pioneering universities is doing the opposite, pursuing deeper adoption, and assuming that any policy built on prevention or sanction will not hold. This paper sides with that second view. Obsessing about detection is a dead end, since generated text is increasingly hard to distinguish from human writing, and detectors still misfire too often to be trusted. What universities need instead is a coordinated effort to set clear, course-by-course rules for GenAI use, redesign assessment toward authentic and interdisciplinary assessment that fosters critical thinking and learner autonomy, and build a serious AI-literacy programme that treats students as critical co-creators rather than passive users. The challenge, though, is not only pedagogical. Adoption at university scale also raises organisational, technical, operational, legal and economic questions that have to be solved together. In this context, the Universidad Politécnica de Madrid (UPM) is developing a strategic and sustainable **AI policy and adoption framework** structured around six dimensions, in which AI functions as an enabler of student autonomy and pedagogical innovation rather than as a threat to be policed.

## 1 INTRODUCTION

The release of ChatGPT in late 2022 was a turning point for higher education. In a matter of months, generative artificial intelligence (GenAI) tools became broadly accessible to students, faculty and administrative staff, opening an unprecedented debate about academic integrity, the future of assessment and the very nature of university nature [1], [2]. Institutional responses across the world have varied widely, from outright prohibitions to early experimentation with institutional licences, ad-hoc guidelines and dedicated governance bodies [3], [4].

Despite this variety, a common pattern stands out. Most universities have started from a defensive position, treating GenAI primarily as a threat to academic integrity rather than as an opportunity to rethink teaching and learning [5]. Effort tends to concentrate on training academic staff in basic prompt techniques, piloting AI-detection tools, and issuing rules that explicitly or in practice discourage student use. This stance is reinforced by genuine concerns that students might rely on GenAI to bypass cognitive effort, eroding skills such as writing, reasoning and problem solving [6], [7].

The detect-and-sanction approach is increasingly hard to sustain. AI-text detectors still produce high false-positive rates and tend to flag non-native English speakers and certain writing styles disproportionately as machine-generated [8]. The rapid evolution of large language models (LLMs) is also making generated text increasingly indistinguishable from human writing, so detection becomes a moving target. Restricting access does not actually eliminate use; it just pushes it underground, away from the educational conversation in which it could be most productive [9].

Our argument is that universities have to move beyond detection towards a joined-up redesign of assessment practices and an institutional way of governing GenAI. We propose a strategic and sustainable AI policy and adoption framework organised around six dimensions: (1) pedagogy and curriculum; (2) AI literacy; (3) governance; (4) technical infrastructure & facilities; (5) ethical and legal; and (6) economic and sustainability. These dimensions cannot be tackled in isolation and have to be addressed simultaneously, together with several cross-cutting sub-dimensions such as data governance, cultural change management, and continuous evaluation and improvement. The Universidad Politécnica de Madrid (UPM), one of the largest technical universities in Spain, is developing such a framework, this is, a framework in which AI functions as an enabler of student autonomy and pedagogical innovation, not as an external service to be policed.

The remainder of the paper is structured as follows. Section 2 sets out the methodology used to analyse the international institutional contexts and the UPM's own institutional context, and to design the proposed framework. Section 3 presents the main results, organised along the six dimensions just outlined. Section 4 discusses the implications and outlines the conclusions and future lines of work.

# 2 METHODOLOGY & RESULTS

Our work follows a qualitative thematic-analysis approach that combines institutional benchmarking with participatory design at UPM. The point is not to evaluate a single intervention, but to articulate a coherent institutional **AI policy and adoption framework** that the university can implement, monitor and refine over time. The benchmarking phase has been completed and has allowed us to extend existing AI policy frameworks for university teaching and learning, such as the one proposed by Chan [10]. The participatory-design phase is currently in progress.

## 2.1 Institutional benchmarking

For the benchmarking we analysed the GenAI strategies of a selected set of universities, including pioneering institutions that have moved beyond detection towards integrated governance and pedagogical redesign. The empirical material comes from a corpus of 91 university policies on generative AI maintained by the Center of Excellence in Teaching & Learning (CETL) at Western University [4]. The corpus covers nine countries, with a strong United States predominance (US n = 75; Canada n = 5; Australia n = 3; Peru n = 3; and one policy each from the United Kingdom, Belgium, Egypt, Singapore, and United Arab Emirates).

***Thematic analysis.*** Using multi-tag, presence/absence coding across 12 themes, we obtained the distribution shown in Fig. 1. The most common themes were AI "*use allowed under certain conditions*" (n = 42), "*training for staff*" (n = 36), and "*decisions usually delegated to the instructor*" (n = 30). Four findings deserve particular attention. First, "*total prohibition*" appears in 14 policies. Second, "*accuracy verification / hallucinations*" appears in only 3% of the corpus (n = 3), even though it is one of the central risks of generative models and a recurrent concern in the technical literature on LLMs. Third, "*data protection / confidentiality*" is addressed in only 14% of policies (n = 13), which is a worrying gap in jurisdictions with demanding legal frameworks (FERPA in the United States; GDPR in Europe). Forth, 26% of the policies mention "*ChatGPT*" by name, which suggests a normative-language bias toward a single commercial product and raises questions about how these policies will age as the ecosystem evolves to include Claude, Gemini, Copilot and a growing set of open-source models.

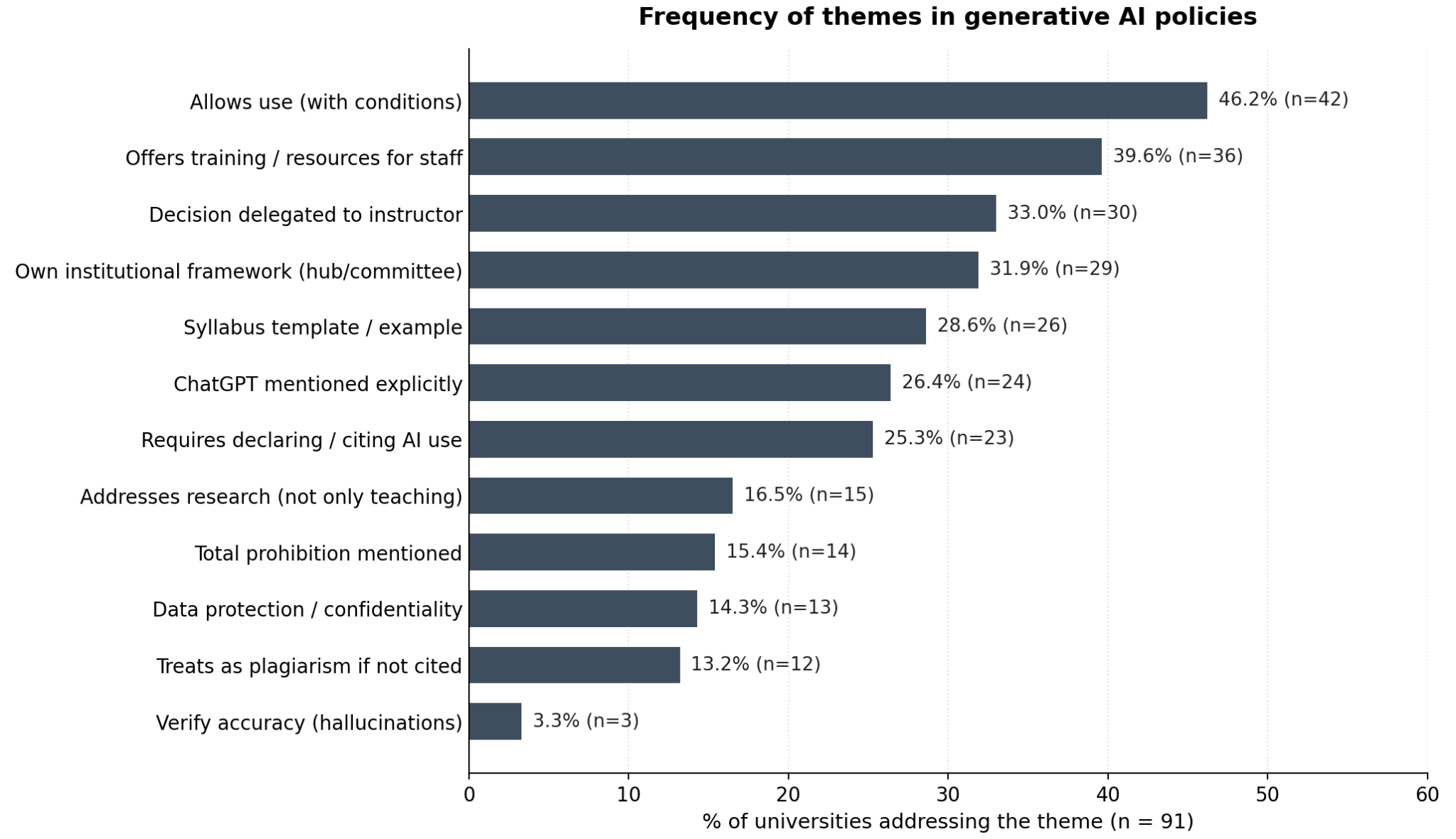

Fig. 1. Frequency of the 12 coded themes (n = 91).

***Four-cluster typology.*** We then assigned each university to a single, mutually exclusive cluster (the one that best described its overall policy stance), which produced four groups. The largest cluster, with 32% of the corpus (n = 29), contains universities that have defined an "*own institutional framework*", i.e, an AI hub, committee, task force or proprietary framework, without that necessarily implying a more restrictive or more permissive stance. The second cluster (28%, n = 25) includes institutions that explicitly avoid taking a substantive position and pass the responsibility to the individual instructor, what we label "*delegated to faculty*". A third cluster (18%) follows a "*permissive with attribution*" line, i.e., it allows GenAI use as long as it is cited, declared or acknowledged as a human-AI co-authorship. A fourth, smaller cluster (11%, n = 10) takes a "*restrictive*" line, where the institution either explicitly includes a 'no AI' option or directly characterises unauthorised GenAI use as plagiarism or academic misconduct. A residual 8.8% (n = 8) could not be classified.

***Analysis using the Chan framework.*** Using Chan's AI education policy framework [10] and its three dimensions, i.e., pedagogical (P), operational (O), and governance (G), the resulting distribution (see Fig. 2) shows a marked pedagogical concentration. Hence, 73% of the policies explicitly address the pedagogical dimension (n = 66), 56% include governance elements (n = 51), and only 16% mention the operational dimension (n = 15). When dimension combinations are examined, the dominant pattern is "P only" (37%), followed by "P+G" (30%). Only four institutions (4%) cover all three dimensions in an integrated way. This suggests that university policies have developed mainly as curricular and academic-integrity responses, with a clear lag in articulating the operational dimension, which involves budgetary, economic and sustainability issues, technical infrastructure, facilities, and institutional deployment, among others.

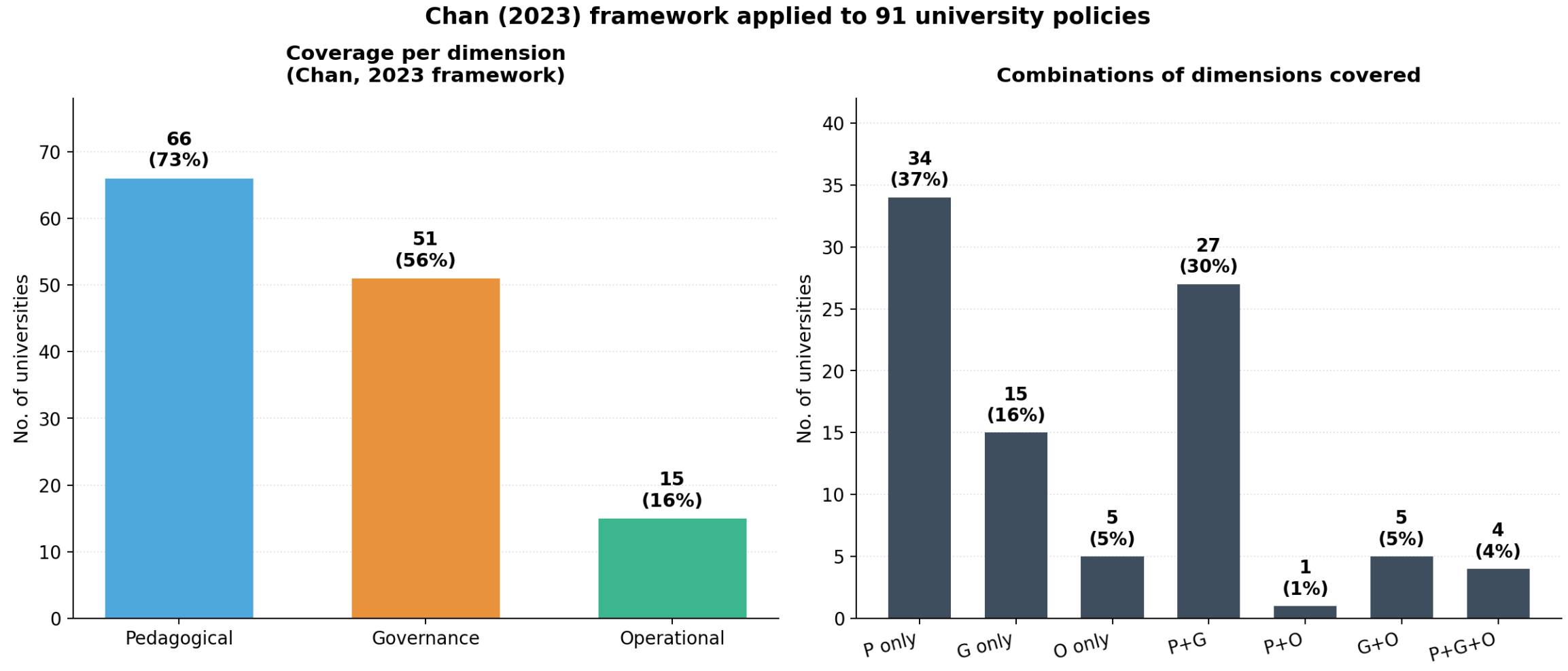

Fig. 2. Distribution of the corpus according to Chan (2023) analytical framework. Left: individual coverage of each dimension. Right: combinations of covered dimensions.

## 2.2 Participatory design at UPM

Within UPM, we are running a series of working groups that bring together teaching staff, researchers, IT services, financial services including invoicing, contracts and procurement units, the data protection office and student representatives. These groups are co-designing the main building blocks of the AI policy and adoption framework and discussing concrete use cases, surfacing pain points: licensing costs, payments, framework agreements, data residency, integration with the institutional systems (such as identity, learning management systems, and office, collaborative, and conferencing tools), among others. This participatory phase has helped us confirm that the challenge is not only (1) pedagogical and curricular: it also involves (2) AI literacy, (3) governance, (4) technical infrastructure and facilities, (5) ethical and legal aspects, and (6) economic and sustainability concerns. The AI policy and adoption framework must articulate these six dimensions in an integrated way (see Fig. 3).

*AI policy and adoption framework for higher education*

<table>
<tr><td colspan="3">5. Ethical & legal<br>Integrity, transparency, GDPR, AI Act</td><td colspan="3">6. Economic & sustainability<br>Cost, agreements, sovereignty, environmental</td></tr>
<tr><td colspan="2">2. AI literacy<br>Students, faculty, staff, university executive board</td><td colspan="2">3. Governance<br>Policies, licences, change management, evaluation</td><td colspan="2">4. Technical infra. & facilities<br>LLM + SLM, HPC, data, security</td></tr>
<tr><td colspan="6">1. Pedagogical & curricular<br>Course-level rules of use · Authentic, interdisciplinary assessment · Critical thinking & learner autonomy</td></tr>
<tr><td colspan="6">Cross-cutting sub-dimensions<br>Data governance · Cultural change management · Continuous evaluation & improvement</td></tr>
</table>

Fig. 3. Six-dimension framework for university-wide adoption of GenAI at UPM. Dimensions 5–6 are cross-cutting conditions; dimensions 2–4 are institutional enablers; dimension 1 is the core mission.

**1. Pedagogical and curricular dimension** *(the core of the framework: the "what for").* This dimension covers the redesign of learning and assessment in a context of generative AI: defining course-level rules of use (forbidden / allowed with disclosure / required), shifting from memory-based examinations toward authentic and interdisciplinary formats (project work, oral defences, portfolios), and developing metacognitive competences such as critical thinking, autonomy and self-regulation. It is also where decisions are made about which competences must not be delegated to AI.

**2. AI literacy and training dimension** *(human enabler).* Although it is sometimes treated as a sub-section of pedagogy, AI literacy deserves a separate dimension because its audiences and formats differ. It covers literacy for students (as critical co-creators, not just users), for teaching staff (prompt design, didactic integration, assessment), for administrative staff (responsible use in administrative tasks) and for the leadership team (strategic literacy). It includes critical verification of outputs, awareness of bias, understanding of the terms of use of the major tools, and specific training for research groups.

**3. Governance and organisational dimension** *(institutional enabler).* This is about how the university decides, regulates and is held accountable on AI matters. It covers the governance structure (committees, AI offices, clear roles), internal policies, differentiated licensing strategies by user group, tenant administration, mechanisms to monitor consumption (especially in token-based pricing models), procurement processes and, crucially, coordination with other universities and consortia in order to avoid duplication and reduce dependency.

**4. Technical infrastructure and facilities dimension** *(material enabler).* The architecture that supports adoption. This includes the hybrid corporate large language models (LLM) and in-house small language models (SLM) stack on top of high-performance computing (HPC) infrastructure, integration with institutional systems (identity, LMS, storage), interoperability and standards, observability and traceability of use, cybersecurity, and the lifecycle management of models and data (versioning, evaluation, retirement). The make-or-buy decision and technical sovereignty also belong here.

**5. Ethical and legal dimension** *(cross-cutting condition of admissibility)*, which combines two related but distinct logics:

*Ethical aspects:* academic integrity, transparency and traceability of AI use, equity and bias mitigation, responsible use, student autonomy, prevention of cognitive dependency.

*Legal aspects:* data protection, intellectual property and copyright over inputs and outputs, compliance with the European AI Act, contracts and processing clauses with providers, accessibility.

**6. Economic and sustainability dimension** *(cross-cutting condition of viability)*. The capacity to keep the framework running over time has four facets:

*Financial services*: invoicing, contracts and procurement units for vendor management, payment workflows, framework agreements, and the administrative capacity needed to evaluate, acquire and sustain AI services under public-sector purchasing and compliance requirements.

*Economic sustainability:* budget, funding models, cost control under metered (token-based) consumption, cost-benefit analysis of make-or-buy, avoidance of prohibitive financial barriers for public universities.

*Strategic sustainability:* reduction of technological dependency on a few large providers, digital sovereignty.

*Environmental sustainability:* energy and carbon footprint of training and inference, especially relevant for in-house HPC infrastructure.

*Social sustainability:* equity of access across user groups (e.g., students with vs. without Plus licences, usage gap).

## 3 PRELIMINAR IMPLEMENTATION

The combined analysis produces several interrelated sets of results that together define the proposed AI policy & adoption framework. UPM is currently implementing this framework through actions that span the six dimensions, which is being promoted by an institutional hub: the OAD (*Oficina de Apoyo a la Docencia*).

**1. *Assessment redesign*.** Through the *Education Innovation Projects* promoted by the OAD, teaching staff are redesigning assessment practices. UPM's approach therefore shifts the focus from detection to assessment redesign. Course-specific rules for GenAI use are being defined at the syllabus level, distinguishing tasks where AI use is forbidden, allowed with disclosure, or actively required. These rules are paired with a progressive move away from memory-based examinations and towards authentic, interdisciplinary assessment formats — project-based learning, case studies, oral defences and portfolios — that emphasise metacognitive competences including critical thinking, learner autonomy and student-centred reflection.

**2. *AI literacy programme*.** The OAD is launching a university-wide AI literacy programme through recurring campaigns named "Autumn Tech", "Spring Tech" and so on. The programme trains teaching staff in prompt design, didactic integration of GenAI in active learning, assessment design under AI-rich conditions, ethical and legal concerns, and specific tools and agentic technologies. Faculty in turn train students not only as AI users but as critical co-creators of AI outputs, with explicit instruction on systematic verification, source triangulation, bias awareness and ethical considerations.

**3. *Institutional AI policy framework*.** The same hub is also defining an explicit AI policy for UPM, involving teaching, learning and researching, which is building on the benchmarking and policy frameworks discussed in Section 2. From a **governance** perspective, deploying GenAI services across a large public university requires explicit decisions on tenant administration, security and compliance management, differentiated licensing strategies, and mechanisms to monitor token-based or metered consumption. Without such mechanisms, costs can escalate quickly and unpredictably, particularly when API-based pay-per-use models are adopted at scale.

**4. Technical infrastructure.** Within this context, UPM is putting in place a hybrid architecture that combines corporate LLMs with in-house developments built on SLMs. The approach builds on the university's experience in HPC and aims to balance cost, data privacy and pedagogical innovation. The architecture is designed to have three complementary layers (not all implemented yet): (i) a curated set of corporate LLM services, accessed through institutional agreements with explicit data-processing clauses, for general productivity tasks (currently Microsoft Copilot Chat and M365 Copilot and Google AI Pro Education) ; (ii) a layer of in-house SLM-based services, deployed on UPM's HPC infrastructure and fine-tuned on selected institutional corpora, for tasks involving sensitive data or where cost predictability is critical; and (iii) a governance and observability layer that monitors usage, costs and compliance across both layers. This hybrid approach is consistent with recent evidence that smaller, well-tuned models can match or approach the performance of larger general-purpose LLMs on domain-specific tasks while offering substantially better control over data, cost and latency. Combined with the assessment redesign and AI-literacy programme described above, it lets UPM treat GenAI as an enabler of student autonomy and educational innovation rather than as an external service whose adoption is dictated by commercial vendors.

**5. *Data protection*.** Data protection is one of the most sensitive issues in the framework. The university community needs literacy in the terms of use of the major commercial AI tools and the ability to judge when commercial solutions from large providers are appropriate, and when locally hosted alternatives based on smaller models are preferable. Sensitive research data, student personal information and confidential administrative documents cannot be uniformly entrusted to external providers without a clear legal and contractual framework aligned with the General Data Protection Regulation (GDPR) and applicable Spanish legislation.

**6. *Economic and sustainability strategy*.** On the economic side, the market shows that providing AI solutions to the entire university community demands significant economic, technical and human resources. Scaling Pro/Plus licences (for example, Microsoft 365 Copilot, ChatGPT Plus, Gemini Advanced or Claude Pro) to all academic staff and students would mean prohibitive financial barriers for a public university, increase technological dependency on a small number of large corporations and add substantial administrative overhead. Our analysis confirms that a uniform commercial-licence strategy is not viable in the medium term and that differentiated, role-based access combined with institutional alternatives is needed.

# 4 CONCLUSIONS

The institutional response to GenAI in higher education is at a critical point. The dominant reactive stance focused on detection, restriction and concerns about academic integrity is increasingly hard to sustain on technical, pedagogical or ethical grounds. Prohibitive policies do not address the deeper question of how universities should educate critical, autonomous learners in an AI-rich environment.

This paper has argued that universities need to move beyond detection and adopt a framework that redesigns pedagogical practices and ties them to AI literacy, governance, ethics and law, and technology and infrastructure. UPM is putting such a model into practice through several interrelated lines of action: a redefinition of rules for GenAI use coupled with authentic assessment; an institutional governance framework that addresses licensing, data protection and cost monitoring; and a hybrid LLM/SLM architecture that builds on UPM's HPC experience to combine corporate services with in-house alternatives.

Several lines of future work follow from this analysis. First, the proposed model needs to be progressively piloted and evaluated, with both quantitative and qualitative indicators of its impact on learning outcomes, faculty workload and operational costs. Second, the AI literacy programme should be articulated as a transversal competence integrated across degree programmes, rather than as a series of isolated workshops and seminars. Third, the hybrid technical architecture should be developed in close coordination with other Spanish and European universities, both to share infrastructure costs and to consolidate sovereign capabilities in AI for education. The goal is not to keep students from using GenAI, but to make sure they learn to use it critically, ethically and creatively, and that the institution itself adopts these tools in a way that is consistent with its public mission.

# ACKNOWLEDGEMENTS

We acknowledge the use of generative AI tools (specifically, large language model assistants) for limited language editing and stylistic review of the manuscript. All conceptual contributions, design decisions, analyses and conclusions presented here are the sole responsibility of the authors, who reviewed and validated every section of the final text.